\documentclass[conference]{IEEEtran}
\IEEEoverridecommandlockouts

\usepackage{amsmath,amssymb,amsfonts}
\usepackage{algorithmic}
\usepackage{graphicx}
\usepackage{textcomp}
\usepackage{xcolor}
\usepackage{booktabs}
\usepackage[numbers]{natbib}
\usepackage{xcolor}
\usepackage{pifont}
\usepackage{multirow}
\usepackage{todonotes}
\usepackage{hyperref}
\usepackage{url}
\usepackage{placeins}
\usepackage{orcidlink}
\usepackage{subcaption}
\usepackage{bm} 
\usepackage{balance}

\usepackage{tikz}
\usetikzlibrary{positioning, shapes.geometric, arrows.meta, fit, backgrounds, calc}
\usepackage{graphicx}

\usepackage{siunitx}
\newcommand{\cmark}{\textcolor{green!70!black}{\ding{51}}}   % ✓
\newcommand{\xmark}{\textcolor{red}{\ding{55}}}             % ✗
\newcommand{\pmark}{\textcolor{orange!90!black}{\ding{72}}} % ~

\def\BibTeX{{\rm B\kern-.05em{\sc i\kern-.025em b}\kern-.08em
    T\kern-.1667em\lower.7ex\hbox{E}\kern-.125emX}}
\begin{document}

\title{Sampling Matters: The Effect of ECG Frequency on Deep Learning-Based Atrial Fibrillation Detection \\
\footnotesize{If accepted, we will purchase an additional page.}
}

\author{
    \IEEEauthorblockN{
        Arjan Mahmuod \orcidlink{0009-0006-4416-8130}\IEEEauthorrefmark{1},
        Adrian Rod Hammerstad \orcidlink{0009-0006-2640-8532}\IEEEauthorrefmark{1},
        Muzaffar Yousef \orcidlink{0009-0003-7240-360X}\IEEEauthorrefmark{1}, 
        Yngve Sebastian Heill \orcidlink{0009-0005-5785-6824}\IEEEauthorrefmark{1}, \\
        Jonas L. Isaksen \orcidlink{0000-0003-3227-1131}\IEEEauthorrefmark{2},
        Jørgen K. Kanters \orcidlink{0000-0002-3267-4910}\IEEEauthorrefmark{2},
        Pål Halvorsen \orcidlink{0000-0003-2073-7029}\IEEEauthorrefmark{3}
        Vajira Thambawita \orcidlink{0000-0001-6026-0929}\IEEEauthorrefmark{3}
    }
    \IEEEauthorblockA{
    \IEEEauthorrefmark{1}Oslo Metropolitan University, Norway \qquad
    \IEEEauthorrefmark{2}University of Copenhagen, Denmark \qquad
    \IEEEauthorrefmark{3}SimulaMet, Norway 
        %\vspace{-5pt}
    }
}

\maketitle

\begin{abstract}
Deep learning models for atrial fibrillation (AF) detection are increasingly trained on heterogeneous electrocardiogram (ECG) datasets with varying sampling frequencies, yet the specific consequences of these discrepancies on model performance, calibration, and robustness remain insufficiently characterized. To address this, we conducted a systematic benchmark using 12-lead, 10-second recordings from the PTB-XL dataset, resampled to target frequencies of 62, 100, 250, and 500~Hz, to evaluate a standard 1-D Convolutional Neural Network (CNN) and a hybrid CNN-Long Short-Term Memory (LSTM) architecture under a rigorous patient-safe cross-validation framework. Our analysis reveals that sampling frequency significantly impacts detection metrics in an architecture-dependent manner; the hybrid CNN-LSTM model demonstrated optimal performance and consistent calibration at intermediate frequencies (100–250 Hz), whereas the 1-D CNN baseline exhibited marked degradation in accuracy and sensitivity at 500 Hz, suggesting increased susceptibility to high-frequency noise. We conclude that ECG sampling frequency is a critical, underappreciated factor in arrhythmia detection, and future foundation models must explicitly control for temporal resolution to ensure clinical reliability and reproducibility.
\end{abstract}


\section{Introduction}
Atrial fibrillation (AF) is the most prevalent sustained cardiac arrhythmia and a key risk factor for both stroke and heart failure~\cite{Vinter2024Temporal}. Deep learning (DL) models have shown strong capability in detecting AF directly from raw electrocardiograms (ECGs), often outperforming traditional rule-based methods~\cite{hsieh2020af,ahmed2023cnn,ping2020cnnlstm}. DL-based AF detection has been widely studied using large public datasets such as PhysioNet, PTB-XL, and various CinC challenges~\cite{PhysioNet-ptb-xl-1.0.1,strodthoff2021ptbxlbenchmark}. 

Despite these advances, DL-based ECG research frequently merges datasets collected under heterogeneous acquisition conditions, including substantial variation in sampling frequency~\cite{strodthoff2021ptbxlbenchmark}. In real-world settings, models are typically trained on ECGs at one sampling rate and later applied to signals acquired at another rate after resampling. Most existing studies either fix the sampling frequency or resample signals without reporting the specific effects of this operation. While prior work has demonstrated that ECG frequency influences QRS detection and spectral features, its role in deep neural representations remains largely unexplored. Similarly, models trained on sinus-rhythm ECGs have yielded strong AF prediction performance~\cite{yuan2023deepAFveterans}, but these typically rely on fixed sampling rates and focus primarily on discrimination rather than probabilistic calibration or robustness against resolution changes. Table~\ref{tbl:lit_review} situates our work relative to these key AF detection and ECG deep learning studies, highlighting that the impact of frequency discrepancy on discrimination performance and calibration has not yet been systematically characterized.

\begin{table}[!t]
\centering
\caption{Comparison of Existing AF Detection Studies with our study. 
(\textbf{12L}=12-Lead ECG, \textbf{AF}=AF Detection, \textbf{Freq:}=Frequency Analysis, \textbf{Calid:}=Calibration Analysis, \textbf{Repro:}=Reproducible Code, \cmark= explicitly addressed, \xmark = not addressed, \pmark = partially addressed)}
\resizebox{\columnwidth}{!}{
\begin{tabular}{lcccccc}
\toprule
\textbf{Study} & 
\textbf{12L} & 
\textbf{AF} & 
\textbf{Freq:} & 
\textbf{Calib:} & 
\textbf{Repro:} \\
\midrule
\citet{daydulo2023cardiac} & \xmark & \pmark & \cmark &  \pmark & \xmark \\
\citet{habib2020choosing} & \pmark & \xmark & \cmark &  \xmark & \xmark \\
\citet{katal2023deep} & \xmark & \cmark & \xmark &  \pmark & \xmark \\
\citet{creasy2025electrocardiogram} & \cmark & \cmark &  \cmark & \xmark & \pmark \\
\citet{ansari2024enhancing} & \cmark & \xmark & \cmark &  \xmark & \cmark \\
\citet{yuan2023deepAFveterans} & \cmark & \cmark & \xmark & \cmark & \xmark \\
\citet{ahmed2023cnn} & \xmark & \pmark & \xmark & \xmark & \xmark \\
\citet{hsieh2020af} & \xmark & \cmark & \xmark & \xmark & \xmark \\
\citet{jo2021xai} & \cmark & \cmark & \xmark & \pmark & \xmark \\
\citet{deng2024cnn} & \xmark & \pmark & \xmark & \xmark & \xmark \\
\citet{ping2020cnnlstm} & \xmark & \cmark & \xmark & \xmark & \xmark \\
\citet{strodthoff2021ptbxlbenchmark} & \cmark & \cmark & \xmark & \pmark & \cmark \\
\citet{feyisa2022lightweight} & \cmark & \pmark & \xmark & \xmark & \xmark \\
\citet{zhang2020interpretableecg} & \cmark & \cmark & \xmark & \xmark & \cmark \\

\midrule
\textbf{This work} & \cmark & \cmark & \cmark & \cmark &  \cmark \\
\bottomrule
\end{tabular}
}
\label{tbl:lit_review}
%\vspace{-10pt}
\end{table}

The sampling frequency is a critical parameter in routine ECG acquisition; it determines temporal resolution and shapes how atrial activity and waveform morphology are captured, directly affecting the visibility of P-waves, QRS upstrokes, and high-frequency details~\cite{book_healthcare_application&deeplearning_2025}. From a machine learning standpoint, modifying the sampling frequency simultaneously changes the effective spectral content and the input sequence length provided to the model, which can significantly influence learned representations. To rigorously assess this, it is crucial to enforce patient-level separation to disentangle the effect of sampling frequency from other confounding factors, such as overlapping patients, label leakage, or inconsistent data splits.

In this study, we introduce a controlled experimental benchmark in which the same 12-lead, 10-second ECG recordings (500 Hz) are resampled to multiple target sampling rates and evaluated within a unified, reproducible framework. Our preprocessing pipeline standardizes inputs through signal denoising, resampling to each target frequency, per-lead Z-score normalization, and fixed-length (10~s) segmentation. A key aspect of our design is that all data splits are patient-safe: each patient is assigned to exactly one fold or split, and every segment derived from that patient’s recordings inherits this assignment. Within this controlled design, we study two architectures that reflect common ECG modeling paradigms: (i) a compact CNN1D baseline and (ii) a hybrid CNN–LSTM model that pairs deep convolutional feature extraction with temporal sequence modeling. Training is conducted with uniform hyperparameters across all sampling frequencies, using 5-fold cross-validation and early stopping, so that any observed performance differences can be attributed to sampling frequency rather than experimental noise.

Our main contributions are:

\begin{itemize}
\item To the best of our knowledge, this is the first systematic study to evaluate the effect of ECG sampling frequency on DL-based AF detection, providing a reproducible, patient-safe resampling benchmark using identical 12-lead, 10-second recordings represented at several target sampling rates.
\item A controlled evaluation of two representative architectures (CNN1D and a hybrid CNN–LSTM) under matched training protocols across sampling frequencies, enabling model-specific analysis of performance, calibration, and robustness.
\item An open, fully reproducible framework—including preprocessing utilities, training scripts, and stored fold checkpoints—made available through a public GitHub repository\footnote{The complete implementation, preprocessing pipeline, and trained checkpoints are publicly available at: {\url{https://github.com/vlbthambawita/AF_detection_vs_freq}}} to facilitate transparent evaluation and future benchmarking.
\end{itemize}

\section{Experimental Design}

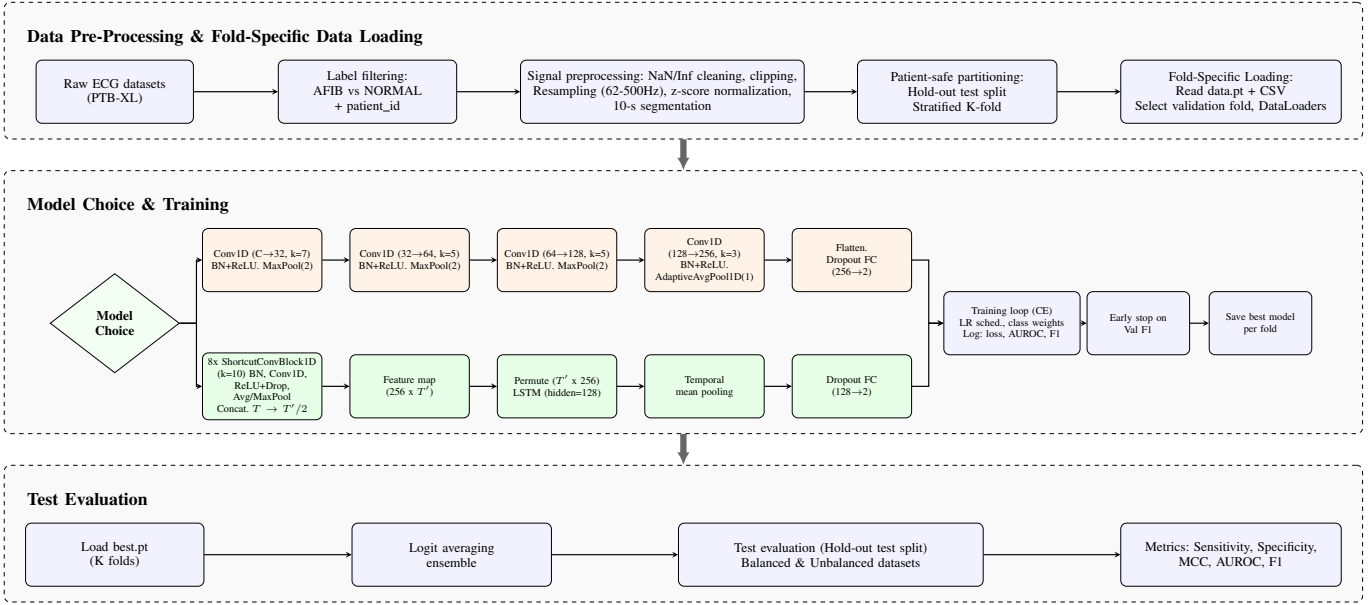
\begin{figure*}[htbp]
\centering
\resizebox{\textwidth}{!}{%
\begin{tikzpicture}[
    >=stealth,
    % General text box styles with expanded widths for Rows 1 & 3
    r1box/.style={draw, rectangle, rounded corners, align=center, fill=blue!5, font=\small, minimum height=1.5cm},
    r3box/.style={draw, rectangle, rounded corners, align=center, fill=blue!5, font=\small, minimum height=1.5cm},
    % Row 2 specific styles
    cnnbox/.style={draw, rectangle, rounded corners, align=center, text width=2.6cm, minimum height=1.5cm, fill=orange!10, font=\scriptsize},
    lstmbox/.style={draw, rectangle, rounded corners, align=center, text width=2.6cm, minimum height=1.5cm, fill=green!10, font=\scriptsize},
    decision/.style={draw, diamond, aspect=1.3, align=center, fill=green!5, font=\small\bfseries, inner sep=2pt},
    r2box/.style={draw, rectangle, rounded corners, align=center, fill=blue!5, font=\scriptsize, minimum height=1.5cm},
    % Bounding box and arrow styles
    rowbox/.style={draw, rectangle, rounded corners, inner sep=0pt, dashed, fill=gray!5},
    title/.style={font=\large\bfseries, align=left},
    arrow/.style={thick, ->},
    thickarrow/.style={-{Latex[length=3mm, width=3mm]}, line width=1.5mm, draw=black!60}
]

% ================= EXPLICIT BOUNDARY NODES FOR PERFECT SYMMETRY =================
% By defining exact top-left (TL) and bottom-right (BR) corners, we force the 
% dashed row boxes to be uniform and the gaps between them to be perfectly equal.

% Row 1 Box Limits (Y spans 4.5 to 7.5)
\node (TL1) at (-2.5, 7.5) {}; \node (BR1) at (29.5, 4.5) {};
% Row 2 Box Limits (Y spans -2.5 to 3.5)  --> Gap from Row 1 is exactly 1.0 unit
\node (TL2) at (-2.5, 3.5) {}; \node (BR2) at (29.5, -2.5) {};
% Row 3 Box Limits (Y spans -6.5 to -3.5) --> Gap from Row 2 is exactly 1.0 unit
\node (TL3) at (-2.5, -3.5) {}; \node (BR3) at (29.5, -6.5) {};

% ================= ROW 1: PRE-PROCESSING & DATA LOADING =================
\node[title] (title1) at (-2.2, 7.1) [anchor=north west] {Data Pre-Processing \& Fold-Specific Data Loading};

\node[r1box, text width=3.5cm] (raw) at (0, 5.5) {Raw ECG datasets\\(PTB-XL)};
\node[r1box, text width=4cm] (filter) at (6, 5.5) {Label filtering:\\AFIB vs NORMAL\\+ patient\_id};
\node[r1box, text width=6.5cm] (preproc) at (13, 5.5) {Signal preprocessing: NaN/Inf cleaning, clipping,\\Resampling (62-500Hz), z-score normalization,\\10-s segmentation};
\node[r1box, text width=4.5cm] (split) at (20, 5.5) {Patient-safe partitioning:\\Hold-out test split\\Stratified K-fold};
\node[r1box, text width=5cm] (dataload) at (26.5, 5.5) {Fold-Specific Loading:\\Read data.pt + CSV\\Select validation fold, DataLoaders};

\draw[arrow] (raw) -- (filter);
\draw[arrow] (filter) -- (preproc);
\draw[arrow] (preproc) -- (split);
\draw[arrow] (split) -- (dataload);

\begin{scope}[on background layer]
    \node[rowbox, fit=(TL1) (BR1)] (R1) {};
\end{scope}

% ================= ROW 2: MODEL CHOICE & TRAINING =================
\node[title] (title2) at (-2.2, 3.1) [anchor=north west] {Model Choice \& Training};

\node[decision, text width=2cm] (choice) at (0, 0) {Model\\Choice};

% CNN1D Path (Top)
\node[cnnbox] (c1) at (3.5, 1.5) {Conv1D (C$\rightarrow$32, k=7)\\BN+ReLU. MaxPool(2)};
\node[cnnbox] (c2) at (7.0, 1.5) {Conv1D (32$\rightarrow$64, k=5)\\BN+ReLU. MaxPool(2)};
\node[cnnbox] (c3) at (10.5, 1.5) {Conv1D (64$\rightarrow$128, k=5)\\BN+ReLU. MaxPool(2)};
\node[cnnbox] (c4) at (14.0, 1.5) {Conv1D (128$\rightarrow$256, k=3)\\BN+ReLU.\\AdaptiveAvgPool1D(1)};
\node[cnnbox] (c5) at (17.5, 1.5) {Flatten.\\Dropout FC\\(256$\rightarrow$2)};

% CNN-LSTM Path (Bottom)
\node[lstmbox] (l1) at (3.5, -1.5) {8x ShortcutConvBlock1D\\(k=10) BN, Conv1D,\\ReLU+Drop, Avg/MaxPool\\Concat. $T\rightarrow T'/2$};
\node[lstmbox] (l2) at (7.0, -1.5) {Feature map\\(256 x $T'$)};
\node[lstmbox] (l3) at (10.5, -1.5) {Permute ($T'$ x 256)\\LSTM (hidden=128)};
\node[lstmbox] (l4) at (14.0, -1.5) {Temporal\\mean pooling};
\node[lstmbox] (l5) at (17.5, -1.5) {Dropout FC\\(128$\rightarrow$2)};

% Convergence Nodes
\node[r2box, text width=3cm] (train) at (21.3, 0) {Training loop (CE)\\LR sched., class weights\\Log: loss, AUROC, F1};
\node[r2box, text width=2.2cm] (stop) at (24.3, 0) {Early stop on\\Val F1};
\node[r2box, text width=2.2cm] (save) at (27.2, 0) {Save best model\\per fold};

% Row 2 Arrows
\draw[arrow] (choice.east) -- ++(0.4,0) |- (c1.west);
\draw[arrow] (choice.east) -- ++(0.4,0) |- (l1.west);
\draw[arrow] (c1) -- (c2); \draw[arrow] (c2) -- (c3); \draw[arrow] (c3) -- (c4); \draw[arrow] (c4) -- (c5);
\draw[arrow] (l1) -- (l2); \draw[arrow] (l2) -- (l3); \draw[arrow] (l3) -- (l4); \draw[arrow] (l4) -- (l5);
\draw[arrow] (c5.east) -- ++(0.4,0) |- (train.west);
\draw[arrow] (l5.east) -- ++(0.4,0) |- (train.west);
\draw[arrow] (train) -- (stop);
\draw[arrow] (stop) -- (save);

\begin{scope}[on background layer]
    \node[rowbox, fit=(TL2) (BR2)] (R2) {};
\end{scope}

% ================= ROW 3: TEST EVALUATION =================
\node[title] (title3) at (-2.2, -3.9) [anchor=north west] {Test Evaluation};

\node[r3box, text width=4cm] (load) at (0, -5.5) {Load best.pt\\(K folds)};
\node[r3box, text width=4.5cm] (ensemble) at (8, -5.5) {Logit averaging\\ensemble};
\node[r3box, text width=7cm] (eval) at (17, -5.5) {Test evaluation (Hold-out test split)\\Balanced \& Unbalanced datasets};
\node[r3box, text width=5cm] (metrics) at (26.5, -5.5) {Metrics: Sensitivity, Specificity,\\MCC, AUROC, F1};

\draw[arrow] (load) -- (ensemble);
\draw[arrow] (ensemble) -- (eval);
\draw[arrow] (eval) -- (metrics);

\begin{scope}[on background layer]
    \node[rowbox, fit=(TL3) (BR3)] (R3) {};
\end{scope}

% ================= INTER-ROW THICK ARROWS =================
\draw[thickarrow] (R1.south) -- (R2.north);
\draw[thickarrow] (R2.south) -- (R3.north);

\end{tikzpicture}
}
\caption{Expanded left-to-right framework pipeline detailing data preprocessing, expanded CNN1D and CNN-LSTM model architectures, and test evaluation phases. Row boundaries and inter-row spacing are completely uniform.}
\label{fig:expanded_pipeline_uniform}
%\vspace{-10pt}
\end{figure*}

To systematically isolate the effect of sampling rate on atrial fibrillation (AF) detection, we implemented a controlled, reproducible benchmarking pipeline. The overall experimental design is illustrated in Figure~\ref{fig:expanded_pipeline_uniform}, which details the end-to-end workflow from raw signal ingestion to comparative evaluation. The framework enforces strict patient-level separation to prevent data leakage and applies identical preprocessing, balancing, and training protocols across four target frequencies (62 Hz, 100 Hz, 250 Hz, and 500 Hz). By standardizing the deep learning architectures and optimization strategies, we ensure that observed variations in performance—measured through discrimination accuracy, probabilistic calibration, and cross-frequency robustness—can be directly attributed to differences in temporal resolution rather than confounding experimental factors.

\subsection{Data}

We utilized 12-lead, 10-second resting ECG recordings from the open-access PTB-XL clinical dataset~\cite{PhysioNet-ptb-xl-1.0.1, wagner2020ptb}, using the provided expert-annotated rhythm labels. The data processing pipeline consisted of three key stages:

\textbf{Preprocessing and Resampling:} Each ECG recording was resampled from its native sampling rate (500 Hz) to four target frequencies (62 Hz, 100 Hz, 250 Hz, and 500 Hz) using an FFT-based procedure (\texttt{scipy.signal.resample}) to interpolate the signal in the frequency domain. Following resampling, signals underwent per-lead Z-score standardization (zero mean, unit variance) to normalize amplitude scaling prior to model input.

\textbf{Cohort Selection and Quality Control:} We restricted the dataset to recordings labeled as Atrial Fibrillation (AFIB) or Normal Sinus Rhythm (NORM), excluding all other diagnostic categories to establish a clean binary classification task. An automated signal quality assessment pipeline was applied to identify and exclude recordings with flatline signals, dead channels, or excessive high-frequency noise. This process retained 11,005 recordings from 10,076 unique patients, confirming that the vast majority of the dataset maintained high acquisition quality with 12 usable leads.

\textbf{Patient-Safe Splitting and Balancing:} To ensure rigorous evaluation, we employed a patient-safe splitting strategy where no patient appeared in more than one fold or split. The dataset was divided into a training/validation set (70\%, $N=7,053$ patients) and an independent held-out test set (30\%, $N=3,023$ patients). While the initial training set contained 7,730 samples, it was heavily imbalanced. To mitigate this, we applied undersampling to the majority class, resulting in a perfectly balanced dataset of 2,058 samples (1,029 per class) for 5-fold cross-validation. The independent test set was evaluated under two complementary conditions. Performance was evaluated on the naturally imbalanced test distribution to mirror real clinical prevalence and deployment conditions

\subsection{Machine Learning Models}

We evaluated two deep learning architectures representing common paradigms in ECG analysis. Both models ingest raw 12-lead ECG signals, with the input sequence length ($T$) scaling linearly with the sampling frequency ($T = 10 \times f_s$).

\subsubsection{1D-CNN (Baseline)} 

This model serves as a standard convolutional baseline. It consists of stacked 1D convolutional blocks equipped with batch normalization, ReLU activations, and max-pooling layers. Feature extraction is followed by global average pooling and a fully connected classification head.

\subsubsection{Hybrid CNN-LSTM} 

To capture both morphological features and long-term temporal dependencies, we implemented a hybrid architecture. The front-end utilizes a deeper convolutional network with residual (shortcut) connections to extract robust local features. These features are then fed into a single-layer Long Short-Term Memory (LSTM) unit to model temporal dynamics before final classification.

Crucially, the architecture depth and hyperparameters remained identical across all experiments; only the input dimension changed, ensuring that performance differences were driven solely by sampling frequency.

\subsection{Training and Evaluation}

\textbf{Optimization:} All models were trained using the Adam optimizer with a constant learning rate and uniform hyperparameters across frequency settings. We employed early stopping based on validation loss to prevent overfitting.

\textbf{Performance Metrics:} Model discrimination was assessed using the Area Under the Receiver Operating Characteristic curve (AUROC), classification accuracy, and F1-score.

\textbf{Calibration and Robustness:} Beyond discrimination, we evaluated the reliability of predicted probabilities using calibration curves, the Brier score, and Expected Calibration Error (ECE). Validation logits were converted to class probabilities using softmax, and the resulting probabilities were stored for each cross-validation fold. ECE was then computed post-hoc by partitioning predicted probabilities into equally spaced confidence bins and measuring the weighted absolute difference between average confidence and empirical accuracy within each bin.

\textbf{Evaluation protocol:}
Model evaluation was performed using two complementary procedures:
single-model validation and ensemble testing. During evaluation, class
probabilities were obtained by applying a Softmax operation to the model
logits, where the probability of atrial fibrillation corresponds to the
positive class (label = 1) extracted as \mbox{$\mathrm{softmax}(\mathbf{z})[:,1]$}.
For the final testing stage, predictions from all cross-validation folds
were ensembled by averaging logits prior to applying Softmax. The
resulting probabilities were used to compute the Area Under the Receiver
Operating Characteristic Curve (AUROC) and the Expected Calibration
Error (ECE). ECE was computed post-hoc from these probabilities using a
NumPy-based implementation. Predictions were partitioned into equally
spaced confidence bins over the interval $[0,1]$, and calibration error
was calculated as the weighted absolute difference between the empirical
positive rate and the mean predicted probability within each bin.
Probabilities were used directly without thresholding, ensuring that
calibration reflects the true probabilistic behaviour of the model.
%Probability-based metrics were computed in a separate testing step
%(Table~\ref{tab:test_sampling}CR1), independent of performance metrics
%based on discrete class predictions (Table~\ref{tab:val_sampling}CR2).

\section{Results}
%\input{Tables/combined_val_test_single_header_dataset_model}
% Requires: \usepackage{booktabs}, \usepackage{multirow}, \usepackage{xcolor}
\begin{table*}[!t]
\centering
\caption{5-fold cross-validation and final imbalanced test performance across ECG sampling frequencies. Validation metrics are reported as mean $\pm$ standard deviation across folds. ECE $\downarrow$ indicates lower is better. Best overall values per dataset (Validation and Test) are highlighted in \textbf{bold}. Best values for each individual model within a dataset are highlighted in \textcolor{red}{red}.}
\label{tab:val_test_sampling_combined}
\footnotesize
\setlength{\tabcolsep}{3.9pt}
\renewcommand{\arraystretch}{1.10}

\resizebox{\textwidth}{!}{
\begin{tabular}{lllcccccccc}
\toprule
\textbf{Dataset} & \textbf{Model} & \textbf{Frequency} &
\textbf{Accuracy} & \textbf{F1} & \textbf{Precision} &
\textbf{Sensitivity} & \textbf{Specificity} & \textbf{MCC} &
\textbf{AUROC} & \textbf{ECE $\downarrow$} \\
\midrule
\multirow{8}{*}{Validation}  & CNN--LSTM & 62 Hz & $0.9844 \pm 0.0074$ & $0.9844 \pm 0.0075$ & $0.9873 \pm 0.0078$ & $0.9816 \pm 0.0156$ & $0.9873 \pm 0.0076$ & $0.9691 \pm 0.0149$ & $0.9952 \pm 0.0040$ & $0.021 \pm 0.002$ \\
 & CNN--LSTM & 100 Hz & $0.9821 \pm 0.0087$ & $0.9820 \pm 0.0089$ & $0.9855 \pm 0.0103$ & $0.9788 \pm 0.0165$ & $0.9854 \pm 0.0102$ & $0.9644 \pm 0.0173$ & $0.9937 \pm 0.0043$ & $0.028 \pm 0.012$ \\
 & CNN--LSTM & 250 Hz & {\textcolor{red}{$\mathbf{0.9878 \pm 0.0047}$}} & {\textcolor{red}{$\mathbf{0.9877 \pm 0.0048}$}} & {\textcolor{red}{$\mathbf{0.9922 \pm 0.0066}$}} & {\textcolor{red}{$0.9835 \pm 0.0135$}} & {\textcolor{red}{$\mathbf{0.9922 \pm 0.0065}$}} & {\textcolor{red}{$\mathbf{0.9758 \pm 0.0095}$}} & $0.9941 \pm 0.0060$ & {\textcolor{red}{$0.020 \pm 0.006$}} \\
 & CNN--LSTM & 500 Hz & $0.9849 \pm 0.0090$ & $0.9847 \pm 0.0091$ & $0.9911 \pm 0.0038$ & $0.9786 \pm 0.0175$ & $0.9912 \pm 0.0037$ & $0.9701 \pm 0.0180$ & {\textcolor{red}{$0.9963 \pm 0.0034$}} & $0.026 \pm 0.013$ \\
 \cmidrule(lr){2-11}
 & CNN1D & 62 Hz & $0.9859 \pm 0.0052$ & $0.9858 \pm 0.0053$ & {\textcolor{red}{$0.9893 \pm 0.0032$}} & $0.9825 \pm 0.0100$ & {\textcolor{red}{$0.9893 \pm 0.0033$}} & $0.9719 \pm 0.0106$ & $0.9956 \pm 0.0041$ & $0.020 \pm 0.005$ \\
 & CNN1D & 100 Hz & {\textcolor{red}{$0.9874 \pm 0.0079$}} & {\textcolor{red}{$0.9873 \pm 0.0080$}} & {$0.9893 \pm 0.0058$} & {\textcolor{red}{$\mathbf{0.9854 \pm 0.0150}$}} & {$0.9893 \pm 0.0057$} & {\textcolor{red}{$0.9748 \pm 0.0159$}} & {\textcolor{red}{$\mathbf{0.9966 \pm 0.0037}$}} & {\textcolor{red}{$\mathbf{0.018 \pm 0.006}$}} \\
 & CNN1D & 250 Hz & $0.9781 \pm 0.0049$ & $0.9781 \pm 0.0051$ & $0.9740 \pm 0.0073$ & $0.9825 \pm 0.0125$ & $0.9737 \pm 0.0074$ & $0.9564 \pm 0.0102$ & $0.9960 \pm 0.0032$ & $0.027 \pm 0.005$ \\
 & CNN1D & 500 Hz & $0.9519 \pm 0.0132$ & $0.9509 \pm 0.0135$ & $0.9679 \pm 0.0117$ & $0.9349 \pm 0.0241$ & $0.9688 \pm 0.0115$ & $0.9046 \pm 0.0264$ & $0.9857 \pm 0.0048$ & $0.038 \pm 0.010$ \\
 \midrule
\multirow{8}{*}{Test}  & CNN--LSTM & 62 Hz & 0.9884 & 0.9581 & 0.9295 & 0.9581 & 0.9884 & 0.9520 & 0.9983 & 0.015 \\
 & CNN--LSTM & 100 Hz & \textbf{\textcolor{red}{0.9905}} & \textbf{\textcolor{red}{0.9657}} & 0.9417 & \textbf{\textcolor{red}{0.9657}} & 0.9905 & \textbf{\textcolor{red}{0.9606}} & \textbf{\textcolor{red}{0.9984}} & \textbf{\textcolor{red}{0.011}} \\
 & CNN--LSTM & 250 Hz & \textbf{\textcolor{red}{0.9905}} & \textbf{\textcolor{red}{0.9657}} & 0.9417 & \textbf{\textcolor{red}{0.9657}} & 0.9905 & \textbf{\textcolor{red}{0.9606}} & 0.9983 & \textbf{\textcolor{red}{0.011}} \\
 & CNN--LSTM & 500 Hz & 0.9887 & 0.9587 & \textbf{\textcolor{red}{0.9429}} & 0.9587 & \textbf{\textcolor{red}{0.9908}} & 0.9523 & 0.9977 & \textbf{\textcolor{red}{0.011}} \\
  \cmidrule(lr){2-11}
 & CNN1D & 62 Hz & 0.9844 & 0.9445 & 0.9061 & 0.9445 & 0.9841 & 0.9366 & \textcolor{red}{0.9975} & 0.017 \\
 & CNN1D & 100 Hz & \textcolor{red}{0.9860} & \textcolor{red}{0.9495} & \textcolor{red}{0.9191} & \textcolor{red}{0.9495} & \textcolor{red}{0.9866} & \textcolor{red}{0.9420} & 0.9972 & \textcolor{red}{0.015} \\
 & CNN1D & 250 Hz & 0.9811 & 0.9328 & 0.8921 & 0.9328 & 0.9817 & 0.9231 & 0.9969 & 0.027 \\
 & CNN1D & 500 Hz & 0.9707 & 0.8963 & 0.8539 & 0.8963 & 0.9750 & 0.8808 & 0.9899 & 0.038 \\
\bottomrule
\end{tabular}
}
\label{tbl:results}

\end{table*}

We present a comprehensive evaluation of the 1D-CNN and Hybrid CNN-LSTM architectures across four target sampling frequencies (62, 100, 250, and 500\,Hz). Performance is assessed using discrimination metrics (ROC and Precision-Recall curves), probabilistic reliability (Calibration curves), and class-wise error rates (Confusion Matrices).

Table~\ref{tab:val_test_sampling_combined} summarizes the validation and test performance. The test results are obtained from an ensemble composed of the best-performing model from each cross-validation fold. The test set follows a naturally imbalanced distribution, reflecting clinical prevalence. While both models achieve high discrimination, their sensitivity to the sampling rate differs significantly. The \textbf{Hybrid CNN–LSTM} maintains robust performance across all frequencies, with optimal discrimination observed at intermediate frequencies (100–250,Hz), whereas the \textbf{1D–CNN} baseline shows a distinct performance degradation at the highest sampling rate (500,Hz), suggesting increased susceptibility to high-frequency noise when relying solely on convolutional features.

\begin{figure*}[t]
    \centering
    % --- Subfigure (a): Left ---
    \begin{subfigure}[b]{0.48\textwidth}
        \centering
        \includegraphics[width=\linewidth]{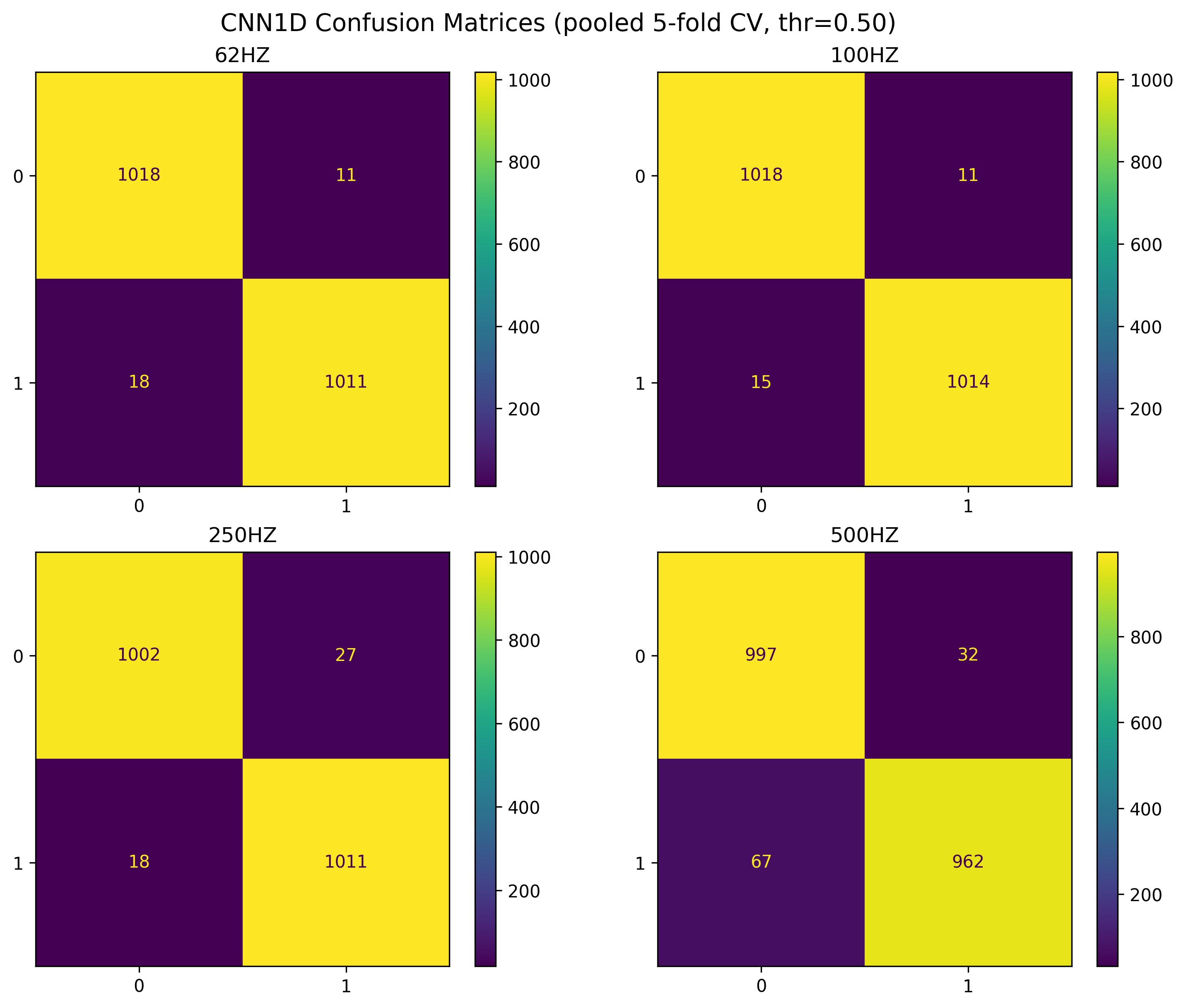}
        \caption{\textbf{1D-CNN} model.}
        \label{fig:confusion_matrices_cnn1d}
    \end{subfigure}
    \hfill % Pushes the figures to the left and right margins
    % --- Subfigure (b): Right ---
    \begin{subfigure}[b]{0.48\textwidth}
        \centering
        \includegraphics[width=\linewidth]{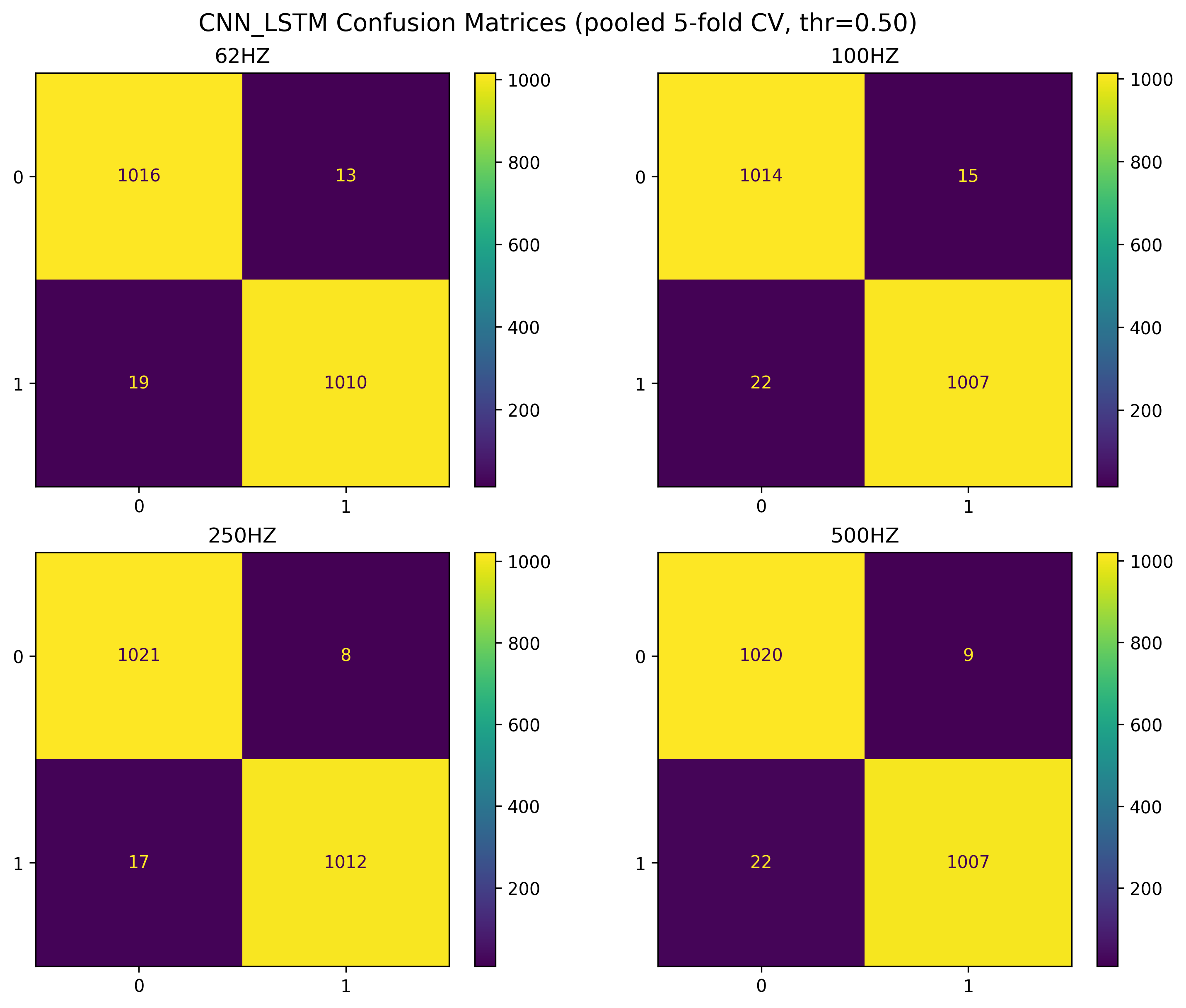}
        \caption{\textbf{Hybrid CNN-LSTM} model.}
        \label{fig:confusion_matrices_cnnlstm}
    \end{subfigure}
    
    \caption{Pooled confusion matrices from 5-fold cross-validation at a decision threshold of 0.5. Left: 1D-CNN baseline showing degradation at high frequency. Right: Hybrid CNN-LSTM showing consistent performance across all frequencies. Note the variation in error counts at 500\,Hz demonstrates stable error rates across the frequency spectrum.}
    \label{fig:confusion_matrices_combined}
    %\vspace{-10pt}
\end{figure*}

Figure~\ref{fig:confusion_matrices_combined} presents the pooled confusion matrices for the 1D-CNN and Hybrid models, respectively. These matrices aggregate True Negatives (TN), False Positives (FP), False Negatives (FN), and True Positives (TP) across all folds at a standard decision threshold of 0.5. Consistent with the metric analysis, the 1D-CNN shows a visible increase in misclassifications at 500\,Hz compared to lower frequencies, whereas the Hybrid model maintains a more stable error distribution.

%\subsection{Discrimination Performance}

\begin{figure*}[t]
    \centering
    \subfloat[1D-CNN Model]{
        \includegraphics[width=0.47\textwidth]{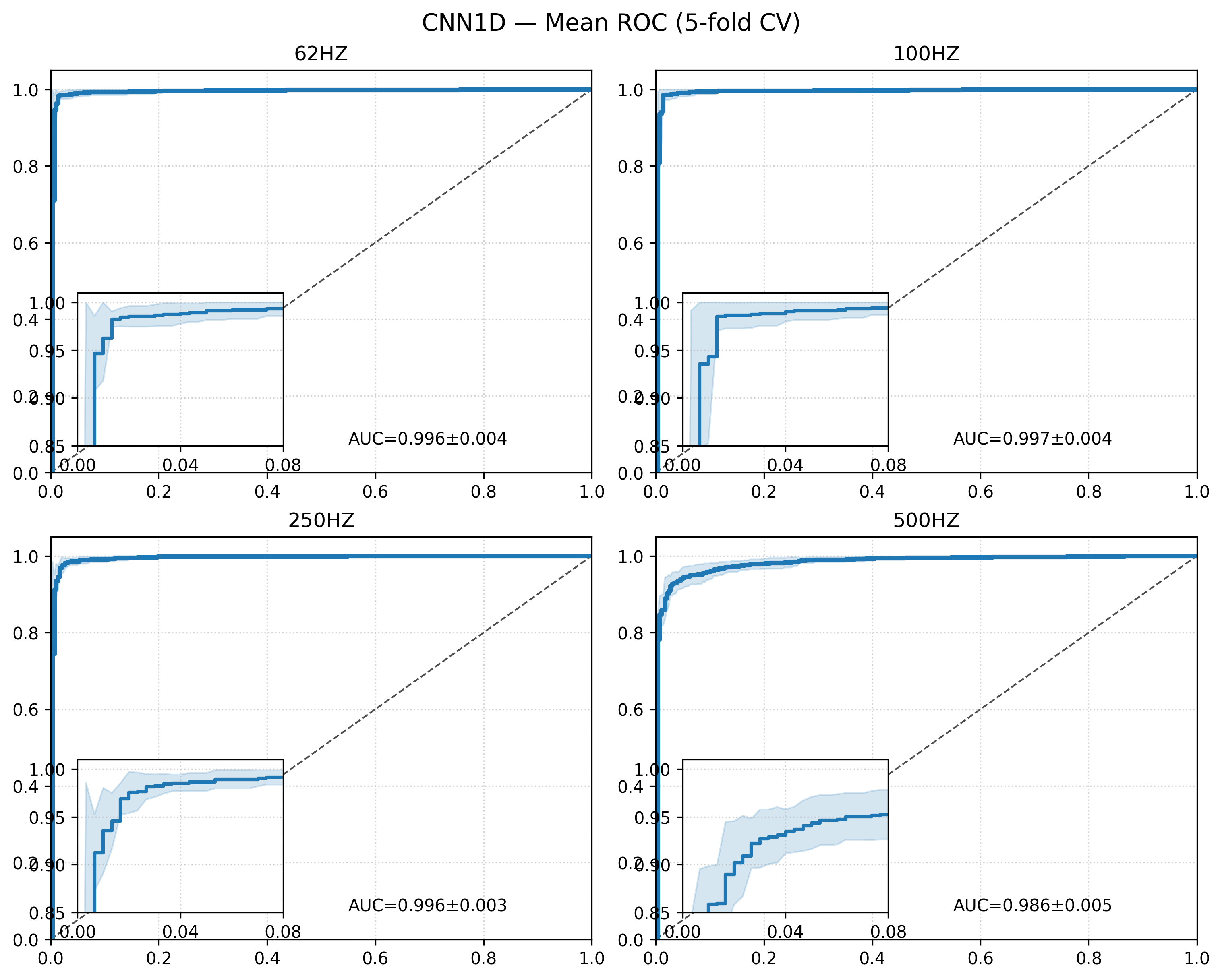}
        \label{fig:roc_cnn1d}
    }
    \hfill
    \subfloat[Hybrid CNN-LSTM Model]{
        \includegraphics[width=0.47\textwidth]{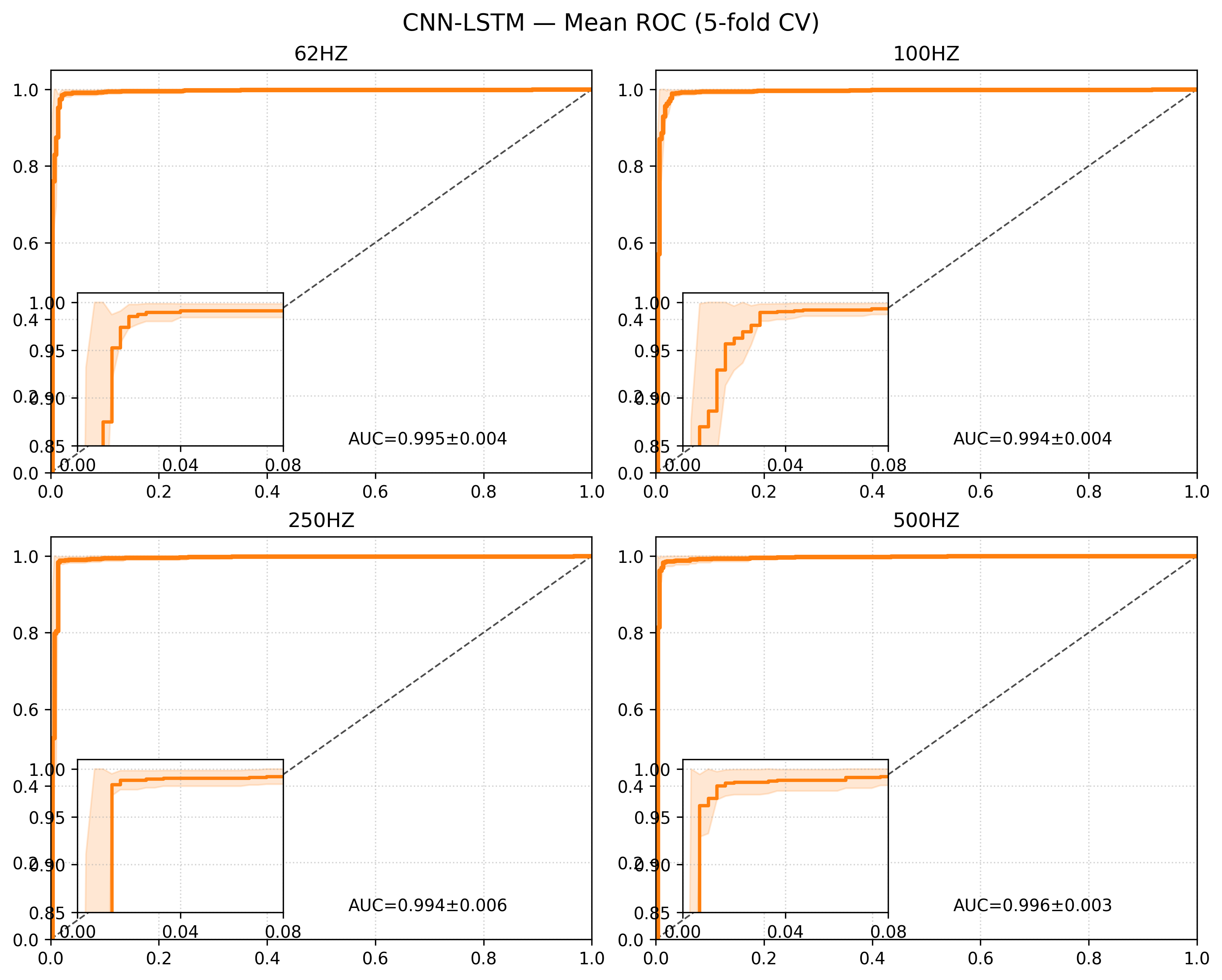}
        \label{fig:roc_cnnlstm}
    }
    \caption{Mean ROC curves across sampling frequencies (62, 100, 250, and 500\,Hz). Shaded regions denote $\pm 1$ standard deviation (x-axis: false positive rate, y-axis: true positive rate).}
    \label{fig:roc_combined}
    %\vspace{-10pt}
\end{figure*}

\begin{figure*}[t]
    \centering
    % --- Left Subfigure: 1D-CNN ---
    \begin{subfigure}[b]{0.48\textwidth}
        \centering
        \includegraphics[width=\linewidth]{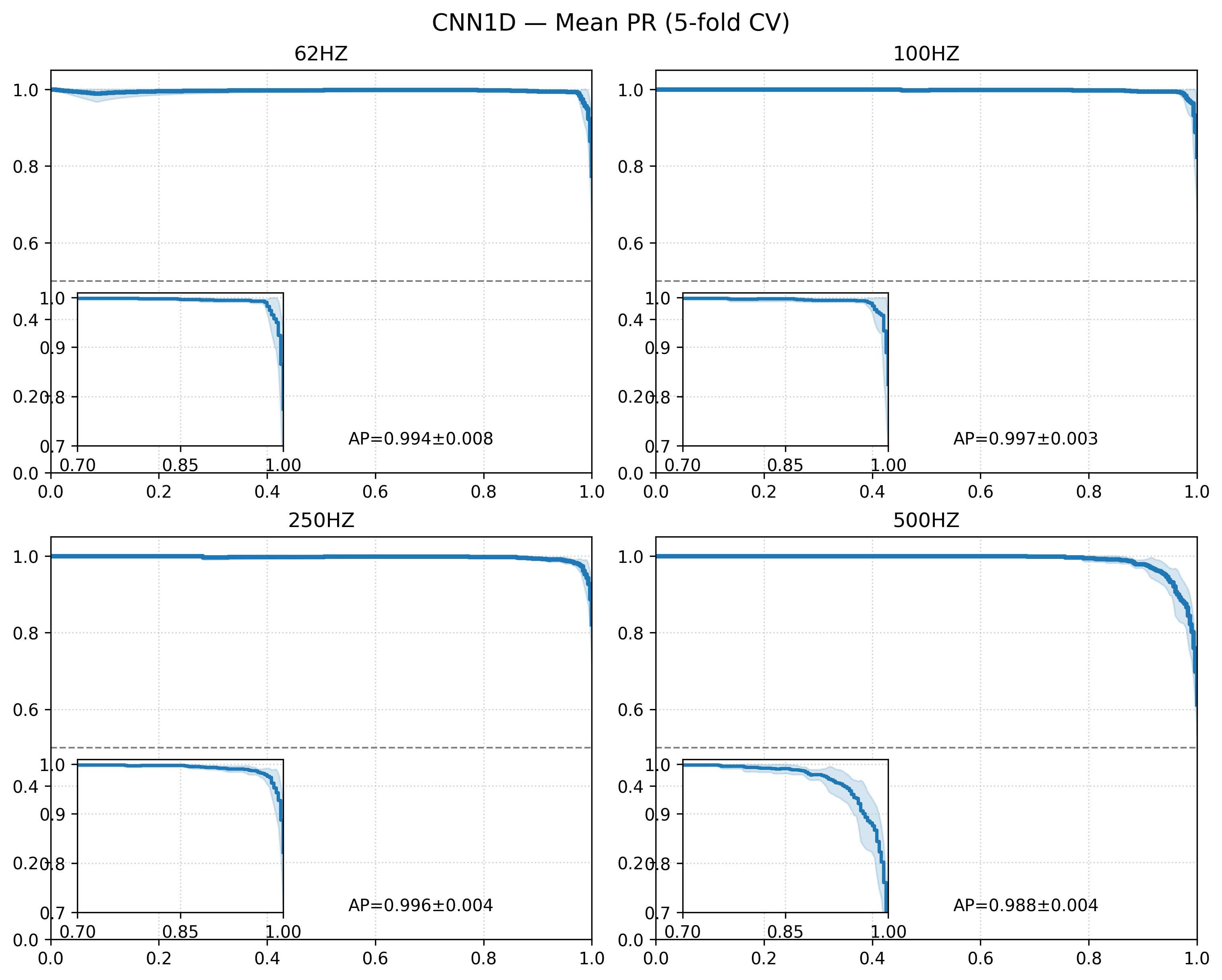}
        \caption{1D-CNN Model}
        \label{fig:pr_cnn1d}
    \end{subfigure}
    \hfill % Spaces them out to left/right margins
    % --- Right Subfigure: CNN-LSTM ---
    \begin{subfigure}[b]{0.48\textwidth}
        \centering
        \includegraphics[width=\linewidth]{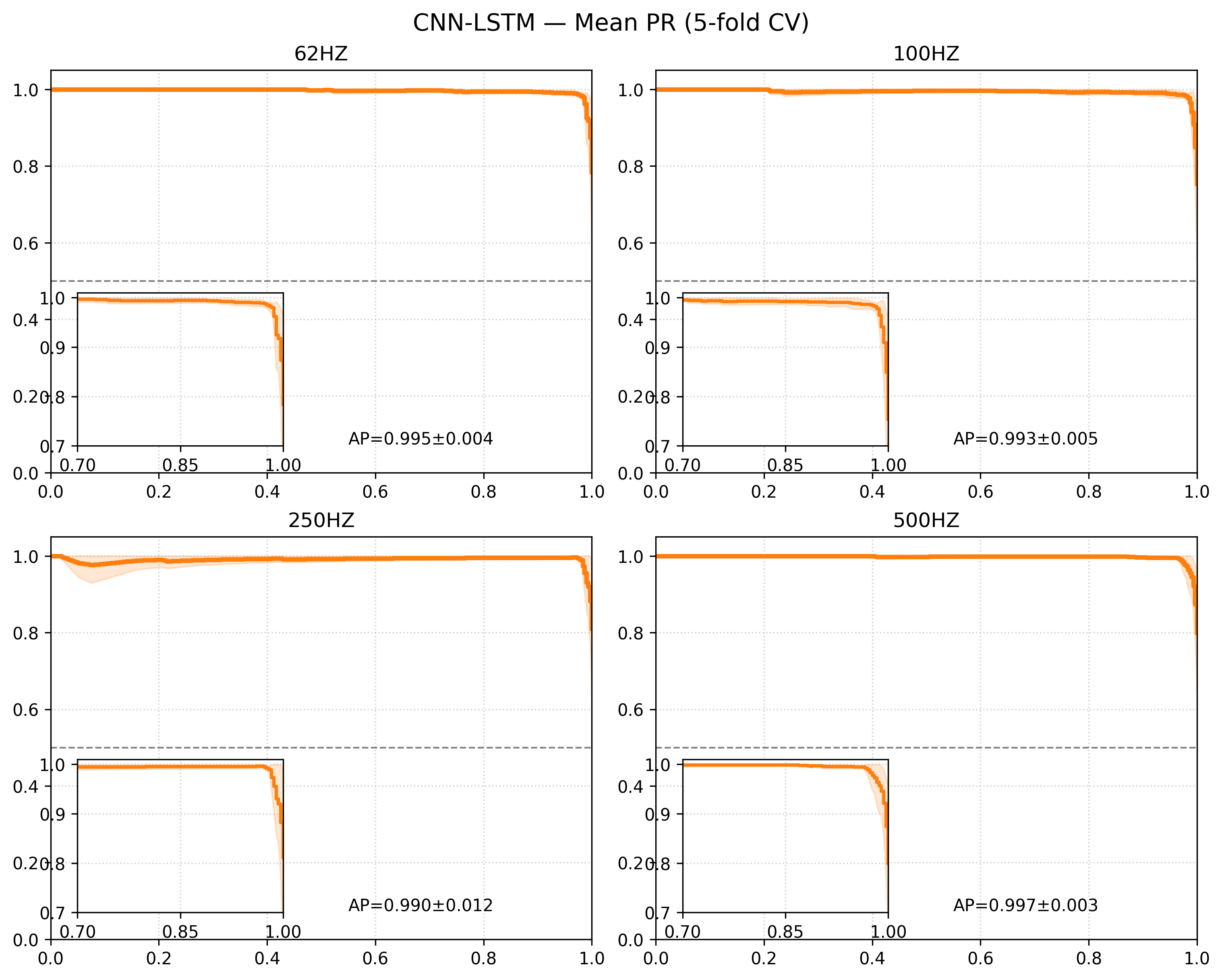}
        \caption{Hybrid CNN-LSTM Model}
        \label{fig:pr_cnnlstm}
    \end{subfigure}
    
    \caption{Mean Precision-Recall (PR) curves across sampling frequencies (62, 100, 250, and 500\,Hz). Curves show the 5-fold cross-validation mean; shaded regions denote $\pm 1$ standard deviation (x-axis: Precision, y-axis: Recall).}
    \label{fig:pr_combined}
    %\vspace{-10pt}
\end{figure*}

Figures~\ref{fig:roc_combined} and \ref{fig:pr_combined} display the mean Receiver Operating Characteristic (ROC) and Precision-Recall (PR) curves for both architectures. Shaded regions denote $\pm1$ standard deviation across the 5 cross-validation folds, illustrating model stability.

In our balanced cross-validation framework ($N=2058$), the prevalence of AF is $\pi = 0.50$. Consequently, the no-skill baseline for the PR analysis is represented by a horizontal line at precision $= 0.50$; curves significantly exceeding this baseline indicate predictive capability superior to random guessing.

% ----------- TABLE Result -----------

%\subsection{Probabilistic Calibration}

\begin{figure*}[!t]
    \centering
    \includegraphics[width=\textwidth]{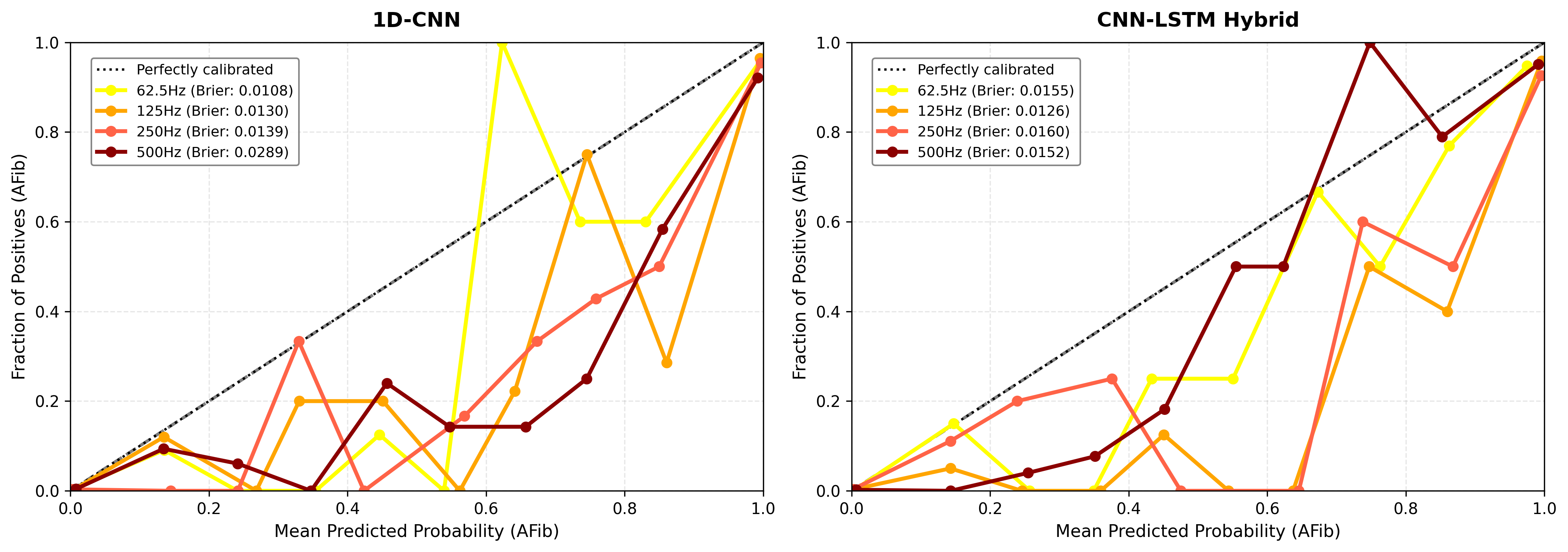}
    \caption{Probability calibration curves for AF detection using CNN1D and CNN--LSTM models across ECG sampling frequencies. The dashed diagonal denotes perfect calibration; curves closer to it indicate better probabilistic calibration.}
    \label{fig:probability_calibration}
    %\vspace{-15pt}
\end{figure*}

To assess the reliability of the predicted risk scores, we analyzed calibration curves (Figure~\ref{fig:probability_calibration}) and Brier scores. Ideally, perfectly calibrated predictions align with the diagonal $y=x$.

\begin{itemize}
    \item \textbf{1D-CNN:} This model exhibits frequency-dependent calibration. It achieves the best alignment (lowest Brier score) at the lowest sampling rate (62\,Hz). However, at 500\,Hz, calibration fidelity degrades, indicating that the model becomes less reliable in its probability estimates as temporal resolution increases.
    \item \textbf{Hybrid CNN-LSTM:} In contrast, the hybrid architecture demonstrates superior robustness. It maintains consistent calibration curves and stable Brier scores across all four frequency settings. This stability suggests that the inclusion of LSTM-based temporal modeling helps regularize the model against variations in sampling rate.
\end{itemize}

These findings imply that while discrimination accuracy may fluctuate, the Hybrid model preserves the clinical reliability of its predicted probabilities even under substantial resampling.

% ----------- FIGURE 1 confustion matix cnn&Hybrid-----------

\section{Discussion}

Our systematic evaluation demonstrates that ECG sampling frequency substantially impacts AF detection performance in an architecture-dependent manner. The \textbf{Hybrid CNN-LSTM} peaks across key metrics at 250\,Hz and maintains strong calibration. Increasing the frequency to 500\,Hz yields no extra benefit, indicating that intermediate resolutions (100--250\,Hz) are sufficient for recurrent models without introducing unnecessary sequence complexity. Conversely, the \textbf{1D-CNN} degrades sharply at 500\,Hz, highlighting its vulnerability to high-frequency noise and expanding dimensionality.

A notable discrepancy occurs at 500\,Hz for the 1D-CNN: AUROC remains high while sensitivity drops. The neural network transforms logits $z$ into posterior probabilities via the softmax function:
$$p_k=\frac{e^{z_k}}{e^{z_0}+e^{z_1}}$$
where $p_1=P(\text{AF})$. Sensitivity relies on a fixed decision threshold of $\tau=0.5$:
$$\hat{y}=\begin{cases}1,&\text{if }p_1\geq0.5,\\0,&\text{otherwise.}\end{cases}$$

This sensitivity drop indicates calibration drift. The predicted probabilities for AF cases shift closer to or below the 0.5 threshold, even as the threshold-independent AUROC correctly ranks positive cases higher. This emphasizes the critical need to report both threshold-dependent and threshold-independent metrics to accurately assess model robustness against sampling rate variations.

\section{Conclusion and Future Work}

This study demonstrates that electrocardiogram sampling frequency is a critical variable that significantly impacts the performance and probabilistic calibration of deep learning models for atrial fibrillation detection. The magnitude of this effect relies heavily on the specific model architecture. A hybrid convolutional and recurrent neural network architecture exhibits superior robustness and maintains consistent calibration across varying frequencies, achieving optimal performance at intermediate temporal resolutions of 100 to 250 Hertz. Purely convolutional baseline models show a marked degradation in sensitivity and accuracy at a high sampling rate of 500 Hertz, which suggests an increased susceptibility to high-frequency noise. Consequently, future foundation models must explicitly control for temporal resolution to ensure clinical reliability and reproducibility.

Building upon our findings , future research should investigate the impact of sampling frequency discrepancies on emerging large-scale ECG foundation models. While our current study evaluated standard convolutional and recurrent architectures, it remains necessary to determine whether pre-trained, transformer-based foundation models inherently learn frequency-invariant representations or if they suffer from similar performance degradation and calibration drift under resolution mismatch. Furthermore, this evaluation must be extended to ECG generative models. As synthetic signal generation—such as DeepFake ECGs ~\cite{thambawita2021deepfake} becomes increasingly prevalent for privacy-preserving data augmentation, it is critical to understand how varying temporal resolutions affect the morphological fidelity, spectral properties, and clinical validity of the synthesized sequences. Ensuring that both discriminative foundation models and generative frameworks explicitly control for temporal resolution will be essential for the safe and reproducible deployment of cardiovascular AI

\section*{Acknowledgment}
%\vspace{-5pt}
%This project is supported by the Innovative Health Initiative Joint Undertaking (IHI JU) and its members under grant agreement 101172997.
This work is part of the European project SEARCH, which is supported by the Innovative Health Initiative Joint Undertaking (IHI JU) under grant agreement No. 101172997. The JU receives support from the European Union’s Horizon Europe research and innovation programme and COCIR, EFPIA, Europa Bio, MedTech Europe, Vaccines Europe, Medical Values GmbH, Corsano Health BV, Syntheticus AG, Maggioli SpA, Motilent Ltd, Ubitech Ltd, Hemex Benelux, Hellenic Healthcare Group, German Oncology Center, Byte Solutions Unlimited, AdaptIT GmbH. Views and opinions expressed are however those of the author(s) only and do not necessarily reflect those of the aforementioned parties. Neither of the aforementioned parties can be held responsible for them. 

\balance
\bibliography{references}

@article{daydulo2023cardiac,
  title={Cardiac arrhythmia detection using deep learning approach and time frequency representation of ECG signals},
  author={Daydulo, Yared Daniel and Thamineni, Bheema Lingaiah and Dawud, Ahmed Ali},
  journal={BMC Medical Informatics and Decision Making},
  volume={23},
  number={1},
  pages={232},
  year={2023},
  publisher={Springer},
  url      = {https://doi.org/10.1186/s12911-023-02326-w}
}

@article{habib2020choosing,
  title={Choosing a sampling frequency for ECG QRS detection using convolutional networks},
  author={Habib, Ahsan and Karmakar, Chandan and Yearwood, John},
  journal={arXiv preprint arXiv:2007.02052},
  year={2020},
  url     = {https://arxiv.org/abs/2007.02052}
}

@article{katal2023deep,
  title={Deep-learning-based arrhythmia detection using ECG signals: A comparative study and performance evaluation},
  author={Katal, Nitish and Gupta, Saurav and Verma, Pankaj and Sharma, Bhisham},
  journal={Diagnostics},
  volume={13},
  number={24},
  pages={3605},
  year={2023},
  publisher={MDPI},
  url     = {https://doi.org/10.3390/diagnostics13243605}
}

@article{creasy2025electrocardiogram,
  title={Electrocardiogram sampling frequency for the optimal performance of complexity analysis and machine learning models: Discrimination between patients with and without paroxysmal atrial fibrillation using sinus rhythm electrocardiograms},
  author={Creasy, Steven and Alexeenko, Vadim and Lip, Gregory YH and Tse, Gary and Aston, Philip J and Jeevaratnam, Kamalan},
  journal={Heart Rhythm O2},
  volume={6},
  number={1},
  pages={48--57},
  year={2025},
  publisher={Elsevier},
  url   ={https://doi.org/10.1016/j.hroo.2024.11.002}
  
}

@article{ansari2024enhancing,
  title={Enhancing ECG-based heart age: impact of acquisition parameters and generalization strategies for varying signal morphologies and corruptions},
  author={Ansari, Mohammed Yusuf and Qaraqe, Marwa and Righetti, Raffaella and Serpedin, Erchin and Qaraqe, Khalid},
  journal={Frontiers in Cardiovascular Medicine},
  volume={11},
  pages={1424585},
  year={2024},
  publisher={Frontiers Media SA},
  url     = {https://doi.org/10.3389/fcvm.2024.1424585}
}

@article{yuan2023deepAFveterans,
  title     = {Deep Learning of Electrocardiograms in Sinus Rhythm From US Veterans to Predict Atrial Fibrillation},
  author    = {Yuan, Neal and Duffy, Grant and Dhruva, Sanket S. and Oesterle, Adam and Pellegrini, Cara N. and Theurer, John and Vali, Marzieh and Heidenreich, Paul A. and Keyhani, Salomeh and Ouyang, David},
  journal   = {JAMA Cardiology},
  volume    = {8},
  number    = {12},
  pages     = {1131--1139},
  year      = {2023},
  publisher = {American Medical Association},
}

@article{PhysioNet-ptb-xl-1.0.1,
  author = {Wagner, Patrick and Strodthoff, Nils and Bousseljot, Ralf-Dieter and Samek, Wojciech and Schaeffter, Tobias},
  title = {{PTB-XL, a large publicly available electrocardiography dataset}},
  journal = {{PhysioNet}},
  year = {2020},
  month = apr,
  note = {Version 1.0.1},
  doi = {10.13026/x4td-x982},
  url = {https://doi.org/10.13026/x4td-x982}
}

@book{book_healthcare_application&deeplearning_2025,
  editor    = {Diwakar, Manoj and Ravi, Vinayakumar and Singh, Prabhishek and Pham, Hoang},
  title     = {Machine Learning and Deep Learning Modeling and Algorithms with Applications in Medical and Health Care},
  series    = {Springer Series in Reliability Engineering},
  publisher = {Springer},
  address   = {Cham},
  year      = {2025},
  edition   = {1},
  isbn      = {978-3-031-98728-1},
  doi       = {10.1007/978-3-031-98728-1}
}

@article{ahmed2023cnn,
  author={Ahmed, Adel A. and Ali, Waleed and Abdullah, Talal A. A. and Malebary, Sharaf J.},
  title={Classifying Cardiac Arrhythmia from ECG Signal Using 1D CNN Deep Learning Model},
  journal={Mathematics},
  volume={11},
  number={3},
  pages={562},
  year={2023},
  url     = {https://doi.org/10.3390/math11030562}
}

@article{hsieh2020af,
  author={Hsieh, Chaur-Heh and Li, Yan-Shuo and Hwang, Bor-Jiunn and Hsiao, Ching-Hua},
  title={Detection of Atrial Fibrillation Using 1D Convolutional Neural Network},
  journal={Sensors},
  volume={20},
  number={7},
  pages={2136},
  year={2020},
  url     = {https://doi.org/10.3390/s20072136}
}

@article{jo2021xai,
  author={Jo, Yong-Yeon and Cho, Younghoon and Lee, Soo Youn and Kwon, Joon-myoung and Kim, Kyung-Hee and Jeon, Ki-Hyun and Cho, Soohyun and Park, Jinsik and Oh, Byung-Hee},
  title={Explainable artificial intelligence to detect atrial fibrillation using electrocardiogram},
  journal={International Journal of Cardiology},
  volume={328},
  pages={104--110},
  year={2021},
  url      = {https://doi.org/10.1016/j.ijcard.2020.11.053}
}

@article{deng2024cnn,
  author={Deng, Jiawen and Yang, Jie and Wang, Xinan and Zhang, Xing},
  title={A Novel Instruction Driven 1-D CNN Processor for ECG Classification},
  journal={Sensors},
  volume={24},
  number={13},
  pages={4376},
  year={2024},
  url     = {https://doi.org/10.3390/s24134376}
}

@article{ping2020cnnlstm,
  author={Ping, Yu and Chen, Chen and Wu, Lin and Wang, Yong and Shu, Ming},
  title={Automatic Detection of Atrial Fibrillation Based on CNN-LSTM and Shortcut Connection},
  journal={Healthcare},
  volume={8},
  number={2},
  pages={139},
  year={2020},
  url     = {https://doi.org/10.3390/healthcare8020139}
}

@article{strodthoff2021ptbxlbenchmark,
  author  = {Strodthoff, Nils and Wagner, Patrick and Schaeffter, Tobias and Samek, Wojciech},
  title   = {Deep Learning for ECG Analysis: Benchmarks and Insights from PTB-XL},
  journal = {IEEE Journal of Biomedical and Health Informatics},
  volume  = {25},
  number  = {5},
  pages   = {1519--1528},
  year    = {2021},
  doi     = {10.1109/JBHI.2020.3022989},
  url     = {https://doi.org/10.1109/JBHI.2020.3022989}
}

@article{feyisa2022lightweight,
  author  = {Feyisa, Desalegn and others},
  title   = {Lightweight Multireceptive Field CNN for 12-Lead ECG Signal Classification},
  journal = {BioMed Research International},
  volume  = {2022},
  pages   = {1--13},
  year    = {2022},
  url     = {https://doi.org/10.1155/2022/4243676}
}

@article{zhang2020interpretableecg,
  author  = {Zhang, Jing and others},
  title   = {Interpretable Deep Learning for Automatic Diagnosis of 12-Lead Electrocardiogram},
  journal = {arXiv preprint arXiv:2010.10328},
  year    = {2020},
  url     = {https://arxiv.org/abs/2010.10328}
}

@article{Vinter2024Temporal,
title={Temporal trends in lifetime risks of atrial fibrillation and its complications between 2000 and 2022: Danish, nationwide, population based cohort study},
author={Nicklas Vinter and P. Cordsen and S. Johnsen and Laila Staerk and Emelia J. Benjamin and Lars Frost and Ludovic Trinquart},
journal={The BMJ},
year={2024},volume={385},
doi={10.1136/bmj-2023-077209}}

@article{wagner2020ptb,
  title={PTB-XL, a large publicly available electrocardiography dataset},
  author={Wagner, Patrick and Strodthoff, Nils and Bousseljot, Ralf-Dieter and Kreiseler, Dieter and Lunze, Fatima I and Samek, Wojciech and Schaeffter, Tobias},
  journal={Scientific data},
  volume={7},
  number={1},
  pages={154},
  year={2020},
  publisher={Nature Publishing Group UK London}
}

@article{thambawita2021deepfake,
  title={DeepFake electrocardiograms using generative adversarial networks are the beginning of the end for privacy issues in medicine},
  author={Thambawita, Vajira and Isaksen, Jonas L and Hicks, Steven A and Ghouse, Jonas and Ahlberg, Gustav and Linneberg, Allan and Grarup, Niels and Ellervik, Christina and Olesen, Morten Salling and Hansen, Torben and others},
  journal={Scientific reports},
  volume={11},
  number={1},
  pages={21896},
  year={2021},
  publisher={Nature Publishing Group UK London}
}
\end{document}